\documentclass[aps,twocolumn,amsmath,amssymb,reprint,numbers,superscriptaddress,noeprint,longbibliography]{revtex4-1}

\usepackage{geometry}
\usepackage{graphicx}
\usepackage{booktabs}
\usepackage{array}
\usepackage{textcase}
\usepackage{relsize}
\usepackage{color}
\usepackage{xcolor}
\usepackage{eqnarray}
\usepackage{bm}
\usepackage{textcomp}
\usepackage{setspace}
\usepackage{braket}
\usepackage{tikz}
\usepackage{hyperref}
\usepackage{newunicodechar}
\usepackage{ulem}[normalem]
\usepackage{makecell}

\usetikzlibrary{positioning}
\hypersetup{
     colorlinks   = true,
     linkcolor    = blue,
     citecolor    = blue,
     urlcolor = blue
}

\definecolor{col2}{rgb}{0.2, 0.5, 1}

\definecolor{col3}{rgb}{0.6, 0.2, 0.6}

\definecolor{col4}{rgb}{1, 0.5, 0.2}

\definecolor{col5}{rgb}{0.2, 0.6, 0.2}

\newunicodechar{μ}{\textmu}

\definecolor{reddy}{rgb}{0.8, 0.1, 0.2}

\definecolor{col6}{rgb}{1, 0.5, 0.5}

\definecolor{col7}{rgb}{0.3, 0.5, 0.5}

\newcommand{\Tone}{$T_1$}
\newcommand{\DoX}{D$^0$X}
\newcommand{\Do}{D$^0$}

\renewcommand{\braket}[3]{\left \langle #1 \left| #2 \right| #3 \right\rangle}

\renewcommand{\ket}[1]{\left| #1 \right\rangle}

\date{\today}
\begin{document}

\title{Ensemble spin relaxation of shallow donor qubits in ZnO}
\author{Vasileios Niaouris}
\email{niaouris@uw.edu}
\affiliation{Department of Physics, University of Washington, Seattle, Washington 98195, USA}
\author{Mikhail V. Durnev}
\affiliation{Ioffe Institute, 194021 St. Petersburg, Russia}
\author{Xiayu Linpeng}
\affiliation{Department of Physics, University of Washington, Seattle, Washington 98195, USA}
\author{Maria L.K. Viitaniemi}
\affiliation{Department of Physics, University of Washington, Seattle, Washington 98195, USA}
\author{Christian Zimmermann}
\affiliation{Department of Physics, University of Washington, Seattle, Washington 98195, USA}
\author{Aswin Vishnuradhan}
\affiliation{Department of Applied Physics and Quantum-Phase Electronics Center, University of Tokyo,
Tokyo, 113-8656, Japan}
\author{Y. Kozuka}
\affiliation{Department of Applied Physics and Quantum-Phase Electronics Center, University of Tokyo,
Tokyo, 113-8656, Japan}
\affiliation{JST, PRESTO, Kawaguchi, Saitama, 332-0012, Japan}
\author{M. Kawasaki}
\affiliation{Department of Applied Physics and Quantum-Phase Electronics Center, University of Tokyo,
Tokyo, 113-8656, Japan}
\author{Kai-Mei C. Fu}
\affiliation{Department of Physics, University of Washington, Seattle, Washington 98195, USA}
\affiliation{Department of Electrical Engineering, University of Washington, Seattle, Washington 98195, USA}


\begin{abstract}

We present an experimental and theoretical study of the longitudinal electron spin relaxation (\Tone) of shallow donors in the direct band-gap semiconductor ZnO. 
\Tone\ is measured via resonant excitation of the Ga donor-bound exciton.
\Tone\ exhibits an inverse-power dependence on magnetic field $T_1\propto B^{-n}$, with $4\leq n\leq 5$, over a field range of 1.75\,T to 7\,T. 
We derive an analytic expression for the donor spin-relaxation rate due to spin-orbit (admixture mechanism) and electron-phonon (piezoelectric) coupling for the wurtzite crystal symmetry. 
Excellent quantitative agreement is found between experiment and theory suggesting the admixture spin-orbit mechanism is the dominant contribution to \Tone\ in the measured magnetic field range. 
Temperature and excitation-energy dependent measurements indicate a donor density dependent interaction may contribute to small deviations between experiment and theory. 
The longest \Tone\ measured is 480~ms at 1.75\,T with increasing \Tone\ at smaller fields theoretically expected. 
This work highlights the extremely long longitudinal spin-relaxation time for ZnO donors due to their small spin-orbit coupling.
\end{abstract}


\maketitle


\section{Introduction}

Shallow impurities in semiconductors are a promising spin-qubit platform for quantum technologies~\cite{ref:morton2011hss,vandersypen2017isq,yamamoto2009ocs}. 
In direct band-gap materials, these spins have an optical interface via the impurity-bound exciton. 
In high-purity crystals, the shallow impurity system can exhibit high optical homogeneity~\cite{ref:fu2005cpt}.
For II-VI semiconductors~\cite{ref:degreve2010pab, linpeng2018cps}
there is also the potential for a nuclear spin-free host with isotope purification, and hence enhanced spin-coherence times~\cite{ref:tyryshkin2012esc, ref:balasubramanian2009usc, tribollet2009ten}.
Within this class of materials, shallow donors (\Do) in ZnO are particularly promising; the donor-bound exciton (\DoX) exhibits narrow inhomogeneous linewidths ($\sim$25\,GHz), short radiative lifetimes ($\sim$1\,ns)~\cite{wagner2011bez}, and a small Huang-Rhys factor ($\sim$0.06)~\cite{wagner2011bez}. 
Additionally, the bound electron exhibits small spin-orbit coupling~\cite{Lew-Yan-Voon:1996tl} which leads to increased isolation from the phonon bath and potential for long longitudinal-spin-relaxation times (\Tone). 
In this paper, we study the dependence of \Tone\ on magnetic field, temperature and excitation energy to gain a fundamental understanding of the mechanisms limiting \Tone\ for shallow donors in ZnO.

This paper is organized as follows:
Sec.~\ref{sec:system} provides an overview of the ZnO donor/donor-bound exciton system and Sec.~\ref{sec:measurement} describes experimental techniques utilized for measuring \Tone. 
Sec.~\ref{sec:exp_field} reports measurements of \Tone\  as a function of magnetic field $\bm{B}$ in both Faraday and Voigt geometries. 
\Tone\ as long as 480\,ms is measured with longer times expected at lower magnetic fields.
In Sec.~\ref{sec:theory} we analytically derive an expression for the \Tone\ dependence on magnetic field and temperature for a single donor, with $T_1\propto B^{-5}$. 
The spin-relaxation model is based on spin-orbit (admixture mechanism) and electron-phonon (piezoelectric) coupling for a wurtzite crystal symmetry. 
In both Faraday and Voigt geometry, remarkable agreement between theory and experiment in the magnitude of \Tone\ is observed. However the experimental exponent is smaller than expected, with the difference more pronounced in the Voigt geometry. 
In Sec.~\ref{sec:exp_energy_temp},
we present measurements of the \Tone\ dependence on the temperature and excitation-energy to further investigate this discrepancy.
We observe dependence of \Tone\ on the excitation energy within the inhomogeneous donor-bound exciton line. 
This variation in \Tone\ at a single field suggests a secondary relaxation mechanism dependent on donor density.
Moreover, the temperature dependence at a given excitation energy is consistent with the expected phonon-occupation model supplemented with an additional excitation-dependent contribution. 
Finally, section~\ref{sec:conclusion} concludes with a brief outlook for the ZnO donor system in the context for quantum information applications.


\begin{figure*}[ht]
    \centering
    \begin{tikzpicture}
        \draw (0, 0) node[inner sep=0] {\includegraphics[width=\textwidth]{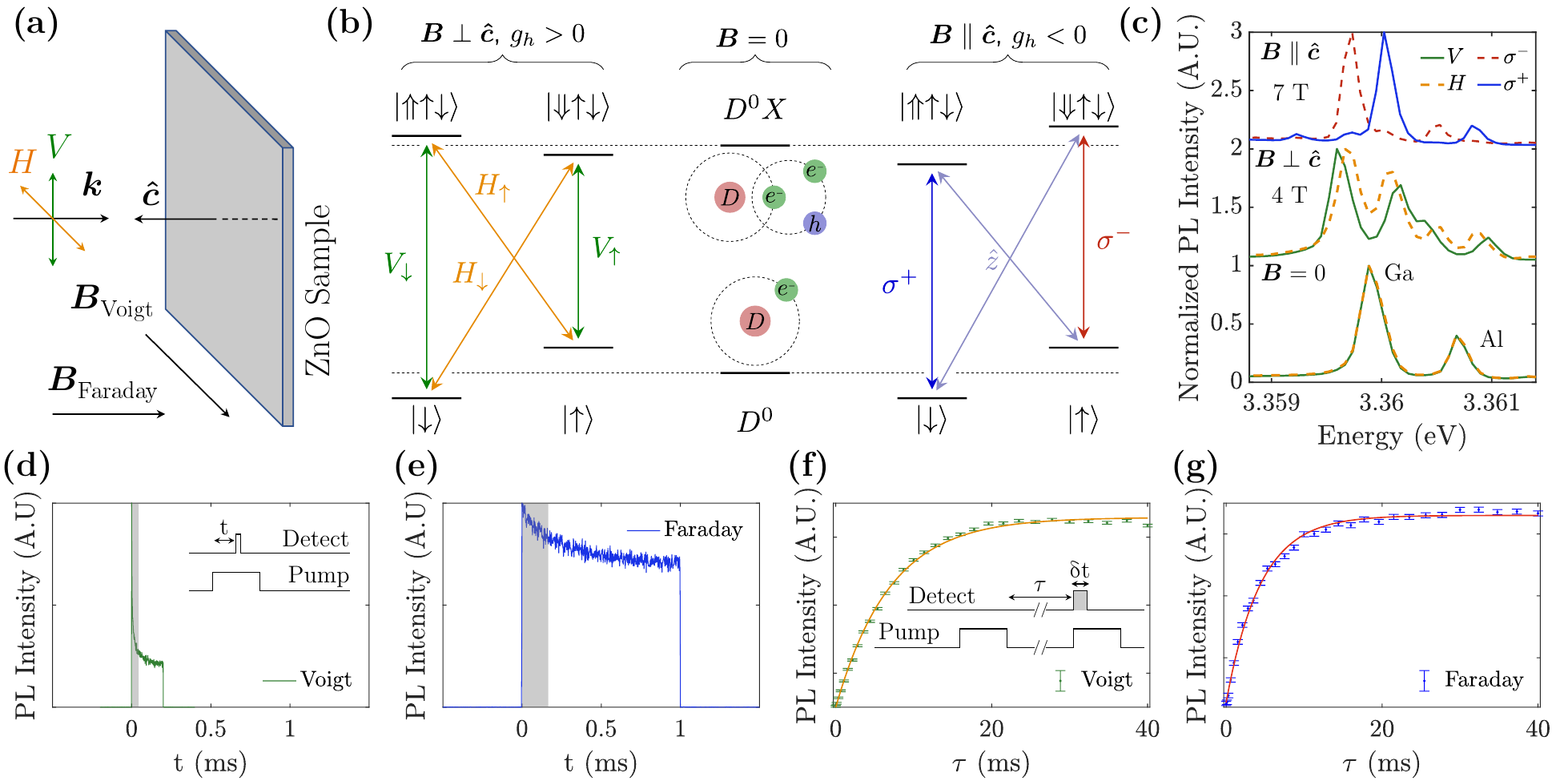}};
    \end{tikzpicture}
    \caption{
    (a) Diagram of sample orientation in experimental setup. H and V are the linear polarization axes of a beam with wavevector $\bm{k}$. The beam propagates parallel to the crystal axis $\bm{\hat{c}}$. The external magnetic field $\bm{B}$ is either parallel ($\bm{B}\parallel\bm{\hat{c}}$) or perpendicular ($\bm{B}\perp\bm{\hat{c}}$) to the crystal axis, labeled as Faraday or Voigt geometry, respectively. 
    (b) Energy diagram of the shallow donor system in Voigt geometry (left), no magnetic field (middle) and Faraday geometry (right). We use green-orange colors for Voigt geometry-related figures, and blue-red colors for Faraday geometry-related figures.
    (c)  PL spectra under 3.45\,eV excitation in the Faraday geometry (7\,T, 1.5\,K), in the Voigt geometry (4\,T, 5.2\,K) and zero field (0\,T, 5.2\,K). 
    (d) Optical pumping curve in the Voigt geometry, 5.5\,T, and 1.5\,K. The inset shows the OP laser sequence, (e) OP curve in the Faraday geometry, 5\,T, and 1.5\,K. (f) Spin-relaxation curve in the Voigt geometry, 5.5\,T, and 1.5\,K. The inset shows the \Tone\ measurement scheme. The error bars depict the photon shot noise. (g) \Tone\ curve in the Faraday geometry, 5\,T, and 1.5\,K.}
    \label{fig:polrulesspectraOPT1}
\end{figure*}

\section{\texorpdfstring{ZnO D\textsuperscript{0}-D\textsuperscript{0}X system}{ZnO D 0-D 0 X system}}
\label{sec:system}
The qubit system studied is the electron spin ($\ket{\uparrow}$ or $\ket{\downarrow}$) of the neutral donor (D$^0$). 
D$^0$ is optically coupled to the donor-bound exciton (D$^0$X), consisting of an electron-hole pair bound to a neutral donor. The D$^0$-D$^0$X transitions form two $\Lambda$-systems consisting of the two electron ground states and an excited state for optical spin manipulation (Fig.~\ref{fig:polrulesspectraOPT1}(b)). 
The Zeeman splitting of the D$^0$ state is determined by the electron spin g-factor.
The Zeeman splitting of the D$^0$X is determined only by the hole g-factor ($\ket{\Uparrow}$ or $\ket{\Downarrow}$), as the two electrons form a spin singlet~\cite{wagner2011bez}. 
In this work, we study a 360\,\textmu m-thick single-crystal ZnO substrate from Tokyo Denpa which is further described in our prior work~\cite{linpeng2018cps}. 
The total donor concentration, including all donors types, is on the order of $\sim10^{16}$-$10^{17}$\,cm$^{-3}$. 
The sample is mounted in a helium immersion cryostat with a superconducting magnet, either in Voigt ($\bm{B}\perp\bm{\hat{c}}$) or Faraday ($\bm{B}\parallel\bm{\hat{c}}$) geometry. Here, $\bm{\hat{c}}$ denotes the [0001] crystal axis and is always parallel to the optical axis $\bm{k}$ (Fig.~\ref{fig:polrulesspectraOPT1}(a)). 
Approximately 10$^6$ donors of all types are in the optical probing volume. 

Figure~\ref{fig:polrulesspectraOPT1}(c) shows the photoluminecence (PL) spectra of the sample in the Faraday and Voigt geometries. At 0\,T, we observe two bright lines at 3.3599\,eV and 3.3607\,eV which closely match the assigned Ga and Al donor transitions~\cite{wagner2011bez}. 
Here, we focus on the Ga donors, with similar behavior expected for the Al donors.

\section{\texorpdfstring{T\textsubscript{1}}{T 1} measurement}
\label{sec:measurement}
\begin{figure}[t]
    \centering
    \includegraphics[width=0.95\linewidth]{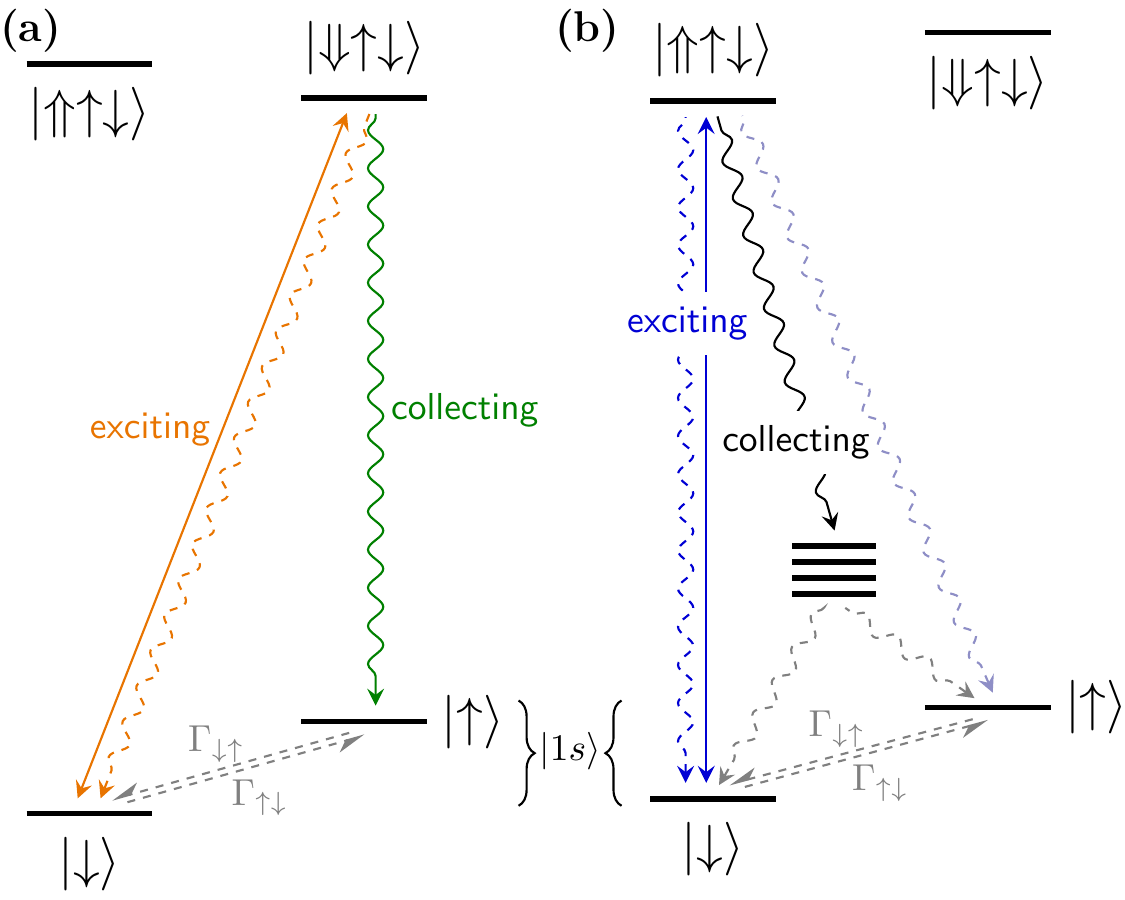}
    \caption{Energy diagram for OP and \Tone\ measurement schemes in the (a) Voigt and (b) Faraday geometry. In (b), the unmarked levels correspond to the energy levels related to the satellite band transitions.}
    \label{fig:EnDiagSceme}
\end{figure}

In the \textit{Voigt geometry} $(\bm{B}\perp \bm{\hat{c}})$, there are four D$^0$-D$^0$X transitions: two with horizontal and two with vertical polarization as shown in  Fig.~\ref{fig:polrulesspectraOPT1}(b). 
These transitions are labeled as 
$H_\downarrow \equiv \ket{\downarrow}\leftrightarrow\ket{\Downarrow\uparrow\downarrow}$, 
$H_\uparrow \equiv \ket{\uparrow}\leftrightarrow\ket{\Uparrow\uparrow\downarrow}$,
$V_\downarrow \equiv \ket{\downarrow}\leftrightarrow\ket{\Uparrow\uparrow\downarrow}$, and
$V_\uparrow \equiv \ket{\uparrow}\leftrightarrow\ket{\Downarrow\uparrow\downarrow}$ with the subscripts corresponding to the ground spin state of the transition.
Prior to measuring the longitudinal spin-relaxation time of the donor ensemble, the spin states are spin-polarized by optical pumping (OP). 
As shown in Fig.~\ref{fig:EnDiagSceme}(a), the spin state is pumped into the $\ket{\uparrow}$ via the $H_{\downarrow}$ transition. 
The measurement signal, collected from the $V_{\uparrow}$ transition, is proportional to the population of the $\ket{\downarrow}$ state. 
Experimentally, we selectively excite the transition of interest via polarized resonant excitation.
Figure~\ref{fig:polrulesspectraOPT1}(d) depicts a typical optical pumping trace in the Voigt geometry. 

We measure the spin-relaxation time by fitting the population recovery of the $\ket{\downarrow}$ state as a function of the time delay $\tau$ between OP pulses (see Fig.~\ref{fig:polrulesspectraOPT1}(f) inset).
The population of $\ket{\downarrow}$ is proportional to the total counts at the start of the OP trace. 
The integration window used is shown in gray in Fig.~\ref{fig:polrulesspectraOPT1}(d). 
Population recovery as a function of delay time $\tau$ is fit with an exponential function, as displayed in Fig.~\ref{fig:polrulesspectraOPT1}(f).

In the \textit{Faraday geometry} $(\bm{B}\parallel \bm{\hat{c}})$, there are also four D$^0$-D$^0$X transitions as shown in Fig.~\ref{fig:polrulesspectraOPT1}(b). 
Two of the transitions are polarized parallel to the optical axis, denoted as $\hat{z}$, and thus cannot be detected. The other two transitions are circularly polarized with
$\sigma^+ \equiv \ket{\downarrow}\leftrightarrow\ket{\Uparrow\uparrow\downarrow}$ and 
$\sigma^- \equiv \ket{\uparrow}\leftrightarrow\ket{\Downarrow\uparrow\downarrow}$.
Because the $\hat{z}$-polarized transitions can not be observed, we utilize the $\sim 10$ times less luminescent two-electron-satellite (TES) transitions (the D$^0$-D$^0$X transitions to the 2s and 2p orbital states of D$^0$), one and two longitudinal optical (LO) phonon replicas (1LO, 2LO), and the first phonon replica TES transitions (1LO-TES), as a probe of the D$^0$ population, as depicted in the energy diagram of Fig.~\ref{fig:EnDiagSceme}(b) and the spectrum in App.~\ref{app:tes_spectrum}. We will denote these transitions as the satellite band transitions.

The OP and \Tone\ measurements in Faraday geometry (Figs.~\ref{fig:polrulesspectraOPT1}(e) and \ref{fig:polrulesspectraOPT1}(g)) are similar to those in Voigt. 
In Faraday geometry, the pump-down time is longer compared to Voigt geometry, because of the $\sim$50-fold weaker $\hat{z}$ dipole transition~\cite{linpeng2020dqd}. Hence, a longer integration window was utilized.

Overall, we have observed a degradation of OP in both Voigt and Faraday geometries with decreasing magnetic field. 
We attribute this behaviour to the 84.8\,μeV (20.5\,GHz) inhomogeneous broadening of the optical transitions, which becomes comparable to the energy difference between the transitions of interest.
In the Faraday geometry, the OP contrast is further degraded by collecting the non-resonant satellite band transitions and at large fields, due to pump-down times comparable to the spin-relaxation time (Fig.~\ref{fig:polrulesspectraOPT1}(e)).

\section{\texorpdfstring{T\textsubscript{1}}{T 1} dependence on magnetic field}
\label{sec:exp_field}

\begin{figure}[!t]
    \centering
    \includegraphics[width=0.976\linewidth]{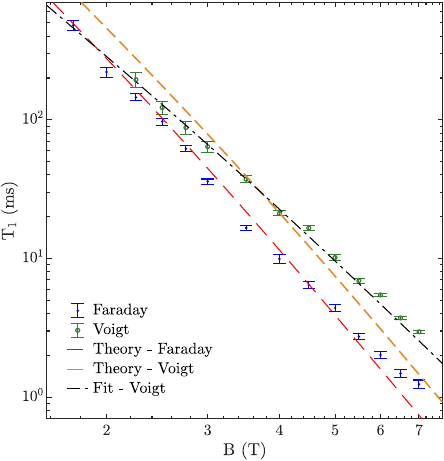}
    \caption{\Tone\ at 1.5\,K as a function of external magnetic field. The error bars correspond to one standard deviation of the \Tone\ fitting error. Theoretical curves are calculated from Eq.~\ref{eq:T1}. The curve fitted to the Voigt geometry data was Eq.~\ref{eq:T1}, where $\Gamma_{\downarrow \uparrow} = a B^{4}$, with a single fitted parameter a.}
    \label{fig:Bdep}
\end{figure}

The magnetic field dependence of \Tone\ at 1.5\,K is shown in Fig.~\ref{fig:Bdep}. 
The minimum magnetic field (2.25\,T in the Voigt and 1.75\,T in the Faraday geometry) was limited by the increased measurement time and decreased optical pumping contrast at lower fields.

As discussed further in Sec.~\ref{sec:exp_energy_temp}, \Tone\ exhibits a resonant excitation laser energy dependence. 
To minimize deviations in \Tone\ due to this dependence, all measurements were taken at the excitation energy where the lowest \Tone\ was expected. 
In Faraday geometry, this corresponds to the maximum of the ($\sigma^+$) transition. 
In the Voigt geometry, the energy was chosen to lie between the unresolved $H_{\downarrow}$ and $V_{\downarrow}$ transitions. 

We observed the longest \Tone, 480\,ms, at 1.75\,T in the Faraday geometry.
This is three times higher than the previously reported \Tone\ \cite{linpeng2018cps} where measurements where only performed in Voigt geometry.
In Faraday geometry, measurements at lower fields are possible due to larger hole Zeeman splitting and polarization selectivity of the optical transition. We are able to observe optical pumping contrast at fields as low as 0.3\,T (App.~\ref{app:low_field_OP}); however, \Tone\ measurements were not performed at this field due to the long duty-cycle and low signal contrast. 
A comparison of the experimental magnetic-field data with theory is made in the next section.

\section{Theoretical description of \texorpdfstring{T\textsubscript{1}}{T 1} and comparison to experiments}
\label{sec:theory}
In this section we consider spin-relaxation mechanisms for donor-bound electrons in ZnO and calculate the corresponding $T_1$. We focus on the spin relaxation mediated by the phonon emission/absorption in the presence of spin-orbit coupling (admixture mechanism), which is the dominant spin-relaxation mechanism for III-V quantum dots~\cite{Khaetskii2001, Woods2002} and donor-bound electrons in GaAs, InP and CdTe compounds~\cite{ref:linpeng2016lsr}. Due to spin-orbit coupling the spin sublevels of the ground donor state contain an admixture of the excited sublevels with opposite spin projections. 
As a result, the matrix elements of the spin-independent electron-phonon interaction between the ground spin sublevels become non-zero resulting in the relaxation of electron spin.  The matrix element of this second-order process is given by
\begin{multline}
\label{Msf}
    M_{\downarrow \uparrow} = \sum \limits_e \left[ \frac{\braket{1s\downarrow}{V_{\rm ph}}{e\downarrow} \braket{e \downarrow }{V_{\rm so}}{1s \uparrow}}{E_{1s\uparrow} - E_{e\downarrow}} \right. \\
    \left. + \frac{\braket{1s\downarrow}{V_{\rm so}}{e\uparrow} \braket{e \uparrow }{V_{\rm ph}}{1s \uparrow}}{E_{1s\downarrow} - E_{e\uparrow}} \right]\:.
\end{multline}
Here $\ket{1s}$ is the ground orbital state of the donor-bound electron, $\ket{e}$ denotes the excited orbital states, $E_{1s \uparrow (\downarrow)}$ and $E_{e \uparrow (\downarrow)}$ are the energies of the ground and excited orbital states with $+1/2$ and $-1/2$ spin projections onto magnetic field, $V_{\rm so}$ and $V_{\rm ph}$ are the operators of the spin-orbit and electron-phonon interaction, respectively.

The spin-orbit Hamiltonian for electrons in wurtzite semiconductors contains linear in wave vector $\bm k$ terms~\cite{rashba1960, Bychkov_1984, Lew-Yan-Voon:1996tl}:
\begin{equation}
\label{so}
    V_{\rm so} = \alpha(\sigma_x k_y - \sigma_y k_x)\:,
\end{equation}
where $\sigma_x$ and $\sigma_y$ are the Pauli matrices, and $\alpha$ is the constant of spin-orbit coupling. 
As for the electron-phonon interaction, we consider only piezoelectric interaction with phonons, since it is more efficient at small phonon wave vectors~\cite{Khaetskii2001,ref:linpeng2016lsr}. The corresponding Hamiltonian is 
\begin{equation}
    V_{\rm ph} = \sqrt{\frac{\hbar}{2\rho \omega_{\bm q, \alpha}}} \exp \left( \mathrm{i} \bm q \cdot \bm r - \mathrm{i} \omega_{\bm q, \alpha} t \right) (e A_{\bm q,\alpha}) b^\dag_{\bm q, \alpha} + \mathrm{c.c.}\:,
\end{equation}
where 
\begin{equation}
    A_{\bm q, \alpha} = \sum \limits_{ijk} \beta_{ijk} \xi_i \xi_j e_k^{(\bm q, \alpha)}\:,
\end{equation}
$\bm q$ and $\alpha$ denote the phonon wave vector and polarization,
$\rho$ is the mass density of the material, $\omega_{\bm q, \alpha}$ is the phonon frequency, $b^\dag_{\bm q, \alpha}$ is the phonon creation operator, $\bm \xi = \bm q/q$ is the unit vector along the phonon wave vector, $\bm e$ is the phonon polarization vector, and $\beta_{ijk}$ is the piezotensor. The nonzero components of $\beta_{ijk}$ in wurtzite media are $\beta_{zxx} = \beta_{zyy} = h_{31}$, $\beta_{zzz} = h_{33}$, $\beta_{xxz} = \beta_{xzx} = \beta_{yyz} = \beta_{yzy} = h_{15}$, where $h_{31}$, $h_{33}$ and $h_{15}$ are piezoelectric constants~\cite{Zook1964}.

In what follows, we use the spherical model for the electronic states of the donor by introducing the averaged electron effective mass $m^*$ and static dielectric constant $\varepsilon$. This model is supported by the small anisotropy of the electron effective mass and dielectric constant in ZnO~\cite{Meyer:2004tt}. Within the spherical approximation, the donor states can be labeled by electron angular momentum $l$ and its projections, in the same way as in the hydrogen atom.
The spin-orbit interaction~\eqref{so} couples the ground $\ket{1s}$ orbital ($l = 0$) and excited $\ket{np}$ orbitals ($l = 1$), where $n = 2,3\dots$. In this section we denote the ZnO $c$-axis as $z$. In the Faraday geometry, when $\bm B \parallel z$, the nonzero matrix elements of $V_{\rm so}$ are $\braket{n p_+\downarrow}{V_{\rm so}}{1s \uparrow}$ and $\braket{1s \downarrow}{V_{\rm so}}{np_- \uparrow}$, where $\ket{p_\pm} = (\ket{p_x} \pm \mathrm{i} \ket{p_y})/\sqrt{2}$. Keeping in mind that the splittings between spin and orbital sublevels induced by magnetic field are much smaller than the energy distance to excited states, as well the relations between the matrix elements  $\braket{n p_+\downarrow}{V_{\rm so}}{1s \uparrow} = -\braket{1s \downarrow}{V_{\rm so}}{np_- \uparrow}$, and $\braket{1s}{V_{\rm ph}}{np_+} = \braket{np_-}{V_{\rm ph}}{1s}$, the spin-flip matrix element~\eqref{Msf} is simplified to
\begin{multline}
\label{MFaraday}
    M_{\downarrow \uparrow} (\bm B \parallel z) = (\hbar \omega_c - 2g\mu_B B) \\
    \times \sum \limits_n \frac{\braket{1s\downarrow}{V_{\rm ph}}{n p_+\downarrow} \braket{np_+ \downarrow }{V_{\rm so}}{1s \uparrow}}{(E_{1s} - E_{np})^2}\:.
\end{multline}
Here $E_{1s}$ and $E_{np}$ are the energies of $1s$ and $np_\pm$ orbitals at zero magnetic field,  $g$ is the electron $g$-factor, $\mu_B$ is the Bohr magneton, and $\omega_c = |e|B/(m^* c)$ is the cyclotron frequency. In the derivation of Eq.~\eqref{MFaraday}, we took into account the splitting $g\mu_B B$ between the spin sublevels of $1s$ and $np$-orbitals, as well as the splitting $\hbar \omega_c$ between the $np_+$ and $np_-$ orbital sublevels. 

In the Voigt geometry, $\bm B \parallel x$, the nonzero matrix elements of $V_{\rm so}$ between the states with opposite spin projections are $\braket{n p_x \downarrow}{V_{\rm so}}{1s \uparrow}$ and $\braket{1s \downarrow}{V_{\rm so}}{n p_x \uparrow}$. Note that here $\uparrow (\downarrow)$ denote the spin projections onto the $x$-axis. Using the same arguments as in the derivation of Eq.~\eqref{MFaraday}, we obtain
\begin{multline}
\label{MVoigt}
    M_{\downarrow \uparrow} (\bm B \parallel x) = - 2g\mu_B B \\
    \times \sum \limits_n \frac{\braket{1s\downarrow}{V_{\rm ph}}{n p_x \downarrow} \braket{np_x \downarrow }{V_{\rm so}}{1s \uparrow}}{(E_{1s} - E_{np})^2}\:.
\end{multline}
In what follows we use the long-wave approximation (LWA) for phonons: $q a_0 \ll 1$, where $q = g \mu_B B/(\hbar s)$, $s$ is the sound velocity, and $a_0$ is the Bohr radius of a donor. This approximation is valid in the whole range of experimentally studied magnetic fields due to a small Bohr radius of shallow donors in ZnO, $a_0 \approx 1.5$~nm~\cite{Meyer:2004tt}. 
 Using LWA, the relation $\braket{np}{\bm k}{1s} = \mathrm{i} m^* (E_{np} - E_{1s}) \braket{np}{\bm r}{1s}/\hbar^2$, and the procedure described in Ref.~\cite{Khaetskii2001}, the matrix elements~\eqref{MFaraday} and \eqref{MVoigt} are simplified to
\begin{eqnarray}
\label{Mfinal}
    M_{\downarrow \uparrow} (\bm B \parallel z) &=& (2g\mu_B B - \hbar \omega_c) \frac{ \alpha m^* \beta (\mathcal E_x + \mathrm i \mathcal E_y) }{2 e \hbar^2} \:, \nonumber \\
    M_{\downarrow \uparrow} (\bm B \parallel x) &=& g\mu_B B \frac{\alpha m^* \beta \mathcal E_x}{e \hbar^2} \:.
\end{eqnarray}
Here $\bm{\mathcal E} = -\mathrm{i} \bm q V_{\rm ph} (\bm r = 0)/e$ is the electric field induced by a phonon at the location of the donor, and 
\begin{equation}
    \beta = 2 e^2 \sum \limits_n \frac{\braket{1s}{x}{np_x}^2}{E_{np} - E_{1s}}
\end{equation}
is the donor polarizability for electric field lying in the $(xy)$-plane. In the spherical approximation that we use, the polarizability is found analytically~\cite{landau3eng}: $\beta = 9 \varepsilon a_0^3/2$.

The spin-flip transition rates are found using Fermi's golden rule, e.g., for the transition from $\ket{1s \uparrow}$ to $\ket{1s \downarrow}$ with emission of a phonon:
\begin{equation}
\label{Gamma0}
    \Gamma_{\downarrow \uparrow} = \frac{2 \pi}{\hbar} \sum \limits_{\bm q, \alpha} \left| M_{\downarrow \uparrow} \right|^2 \delta( \hbar q s_\alpha - g \mu_B B)\:.
\end{equation}
Accurate averaging over $\bm q$ direction in Eq.~\eqref{Gamma0} is difficult due to the complicated phonon structure in wurtzite crystals. However simplified estimations can be made within the model of the effective isotropic elastic medium, when the longitudinal and transverse phonons are decoupled and propagate with isotropic sound velocities $s_l$ and $s_t$~\cite{Zook1964}. This approximations seems reasonable since the relations $c_{11} \approx c_{33}$, $c_{12} \approx c_{13}$ and $c_{44} \approx (c_{11} - c_{12})/2$ hold for the elastic moduli values in ZnO~\cite{madelung1999}. The summation in Eq.~\eqref{Gamma0} is then performed for a longitudinal mode with $\bm e^{(\bm q, l)} = \bm \xi$ and two transverse modes with $\bm e^{(\bm q, t)} \perp \bm \xi$. Averaging over $\bm \xi$ direction for transverse phonons is done with the use of the formula $\langle e_i^{(\bm q, t)} e_j^{(\bm q, t)} \rangle = (\delta_{ij} - \xi_i \xi_j)/2$. By substituting the matrix elements~\eqref{Mfinal} in Eq.~\eqref{Gamma0} and performing the summation, we obtain
\begin{equation}
\label{eq:Gamma_final}
    \Gamma_{\downarrow \uparrow} (\bm B \parallel z) = \frac{\Lambda \Delta_1^3 \Delta_2^2 }{\hbar E_{1s}^4}\:,~~~~
    \Gamma_{\downarrow \uparrow} (\bm B \parallel x) = \frac{\Lambda \Delta_1^5 }{2 \hbar E_{1s}^4}\:,
\end{equation}
where 
\begin{multline}
\label{Lambda}
    \Lambda = \frac{9 (e \alpha)^2}{448 \pi \rho \hbar^3} \left( \frac{5h_{33}^2 + 8h_{31}^2 + 32h_{15}^2}{5 s_l^5} \right. \\
    \left. + \frac{4h_{33}^2 + 4h_{31}^2 + 52 h_{15}^2}{5 s_t^5} \right)\:,
\end{multline}
$\Delta_1 = g \mu_B B$, and $\Delta_2 = \Delta_1 - \hbar \omega_c/2$. Calculation based on Eqs.~\eqref{eq:Gamma_final} and \eqref{Lambda} using parameters listed in Tab.~\ref{table:par} yields $\Lambda \approx 0.02$, $\Gamma_{\downarrow \uparrow} (\bm B \parallel z)/B^5 \approx 0.08$~s$^{-1}$T$^{-5}$, and $\Gamma_{\downarrow \uparrow} (\bm B \parallel x)/B^5 \approx 0.04$~s$^{-1}$T$^{-5}$.

\begin{table*}[!]
\begin{center}
\begin{tabular}{c c c c c c c c c c}
\hline
\hline
 $\rho$ (kg/m$^3$) &  $m^*/m_0$ & $\varepsilon$ & $\alpha$ (meV \AA) & $g$ & $h_{33}$ (V/m) & $h_{31}$ (V/m) & $h_{15}$ (V/m) & $s_l$ (m/s) & $s_t$ (m/s)  \\
\hline  
\\
5.6$\times$10$^3$ & 0.25  & 8.1 & 1.1 &
2 & 1.5$\times$10$^{10}$ & -0.6$\times$10$^{10}$  & -0.6$\times$10$^{10}$ &
6.1$\times$10$^3$ & 2.9$\times$10$^3$ \\
& \cite{Meyer:2004tt} & ~\cite{Meyer:2004tt} & ~\cite{Lew-Yan-Voon:1996tl} & ~\cite{linpeng2018cps} & ~\cite{madelung1999} & ~\cite{madelung1999} & ~\cite{madelung1999} & ~\cite{madelung1999} & ~\cite{madelung1999} \\
\\
\hline
\hline
\end{tabular}
\end{center}
\caption{Parameters of ZnO used in calculations of $T_1$. The piezoelectric constants are calculated using the values of piezoelectric stress moduli $e_{ij}$ as $h_{ij} = e_{ij}/(\varepsilon \varepsilon_0)$, where $\varepsilon$ is the static dielectric constant, and $\varepsilon_0$ is the vacuum permittivity. the electron effective mass and dielectric constant are calculated as $3/m^{*} = 1/m_{e\parallel} + 2/ m_{e\perp}$, $3/\varepsilon = 1/\varepsilon_{\parallel} + 2/ \varepsilon_{\perp}$, the sound velocities $s_l = \sqrt{c_{11}/\rho}$, $s_t = \sqrt{c_{44}/\rho}$.}
\label{table:par}
\end{table*}

The measured spin-relaxation time \Tone\ at nonzero temperature is $T_1 = 1/[\Gamma_{\downarrow \uparrow}(T) + \Gamma_{\uparrow \downarrow}(T)]$, where $\Gamma_{\downarrow \uparrow}(T) = \Gamma_{\downarrow \uparrow} [N_{\rm ph}(T) + 1]$, $\Gamma_{\uparrow \downarrow}(T) = \Gamma_{\downarrow \uparrow} N_{\rm ph} (T)$, and $N_{\rm ph} (T)$ is the phonon occupation number. With that we find 
\begin{equation}
    \label{eq:T1}
    T_1 = \frac{\mathrm{e}^{\gamma} - 1}{\Gamma_{\downarrow \uparrow} (\mathrm{e}^{\gamma} + 1)}\:,
\end{equation}
where $\gamma = g\mu_\mathrm{B} B/k_\mathrm{B} T$ and $k_\mathrm{B}T$  is the thermal energy.



We note that the simple model used here to calculate donor electronic states does not take into account the anisotropy of the electron effective mass and the presence of a short-range impurity potential~\cite{Meyer:2004tt}. 
These effects result in a small shift of $E_{1s}$~\cite{Meyer:2004tt} and consequently slightly affect the \Tone\ value through the denominator in Eq.~\eqref{eq:Gamma_final}. 
However, we neglect these small corrections in order to keep our model simple. 
We also note, that the spin-flip rate~\eqref{eq:Gamma_final} is quite universal, since it does not depend on the electron effective mass, as the $E_{1s}$ value can be taken from experiment. 
Also, other mechanisms of electron-phonon interaction, such as deformation potential and direct spin-phonon interaction are less efficient at small phonon wave vectors, result in smaller spin-flip rates and $T_1 \propto B^{-7}$ dependence not observed in the experiment~\cite{ref:linpeng2016lsr}.

Figure~\ref{fig:Bdep} includes the theoretically expected \Tone\ curves.
The theoretical \Tone\ curves include no fit parameters and lie remarkably close to the experimental values.
The calculated \Tone\ values are sensitive to the values of the piezoelectric constants, which have quite a wide spread in the literature. This spread may result in $\sim$2-times change of the calculated \Tone\, which still gives a good agreement with experiment.
Additionally, through much of the experimental range of magnetic fields, the expected relationship $T_1(\bm B \parallel x) \approx 2 T_1(\bm B \parallel z)$ is approximately observed.
However, the experimental exponential dependence deviates from the expected $B^{-5}$ (see Eqs.~\ref{eq:Gamma_final} and~\ref{eq:T1}). 
Specifically, in the Faraday geometry, a softening of the exponent is observed at higher fields; while, in the Voigt geometry, a $B^{-4}$ dependence is observed across the full experimental range of magnetic fields (as shown in Fig.~\ref{fig:Bdep}). 
If we extrapolate \Tone\ to lower fields, we may expect a cross-over between Faraday and Voigt geometry \Tone\ to occur for fields below 2\,T.
This discrepancy suggests that while spin-orbit coupling is the dominant relaxation mechanism for donor-bound electrons in ZnO, there is an additional mechanism. 

\section{Excitation energy and temperature dependence} 
\label{sec:exp_energy_temp}
\begin{figure}[!t]
    \centering
     \includegraphics[width=0.976\linewidth]{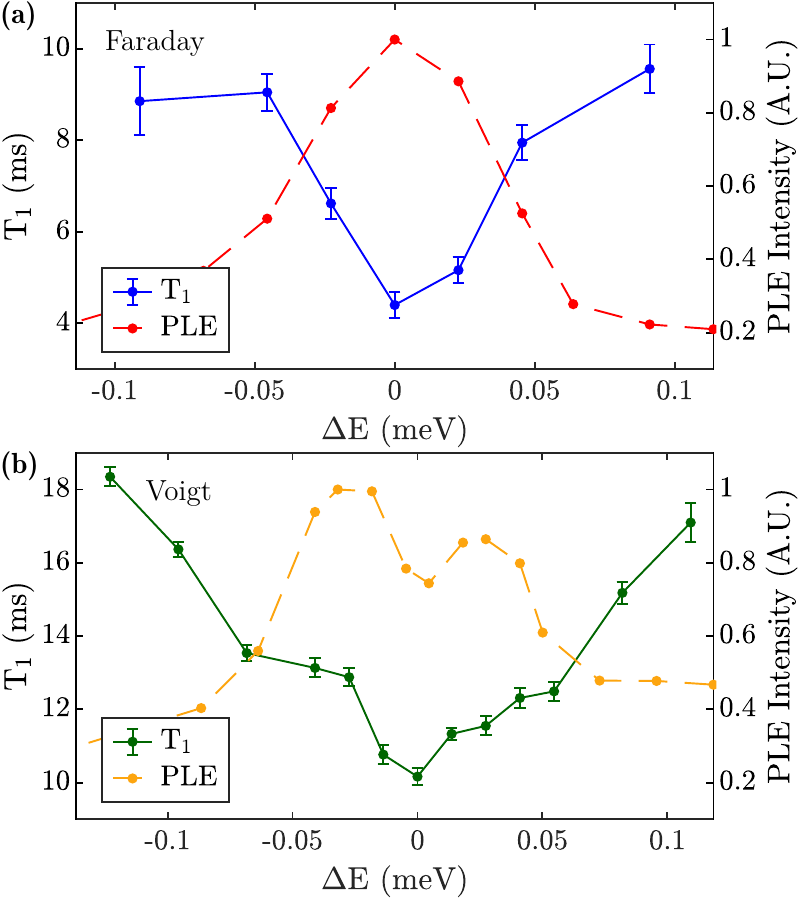}
    \caption{\Tone\ and PLE at 5\,T and 1.5\,K for varying excitation energy detuning $\Delta E$ in (a) Faraday and (b) Voigt geometry. The error bars correspond to one standard deviation of the \Tone\ fitting error.}
    \label{fig:EnDep}
\end{figure}
\Tone\ at a fixed magnetic field and temperature was found to depend on the optical pumping excitation energy.  
Figures~\ref{fig:EnDep}(a,b) show the photoluminescence excitation (PLE) spectra (dashed curves) and \Tone\ (solid curves) in Voigt and Faraday geometry at 5\,T and 1.5\,K. The PLE spectra were taken by tuning the excitation laser over the $H_{\downarrow}$ and $\sigma^+$ transitions respectively, while collecting the satellite band transitions. 
In Faraday geometry, we observe the expected PLE peak. \Tone\ reaches a minimum value near the maximum of the PLE. 
In Voigt geometry, two peaks are observed. The low energy peak corresponds to resonant excitation of the $H_{\downarrow}$ transition (Fig.~\ref{fig:polrulesspectraOPT1}(b)).
The high energy peak corresponds to the resonant excitation of the $V_{\downarrow}$ transition. 
The observation of the high energy peak indicates either a relaxation of the polarization selection rules or an impure polarization excitation. In Voigt geometry, the spin-relaxation time reaches a minimum in between the two peaks. 

A dependence on the pump laser excitation energy for \Tone\ is not expected for an isolated donor, as the excitation pulse is only used for spin initialization and the relaxation process occurs while the excitation pulse is off. 
Laser leakage through the acousto-optic modulator (AOM) could result in optical pumping during the spin recovery period which would be more efficient on-resonance if the resonance line is homogeneously broadened. 
This potential cause of a reduced \Tone\ on-resonance can be ruled out due to the high AOM extinction ratio ($>$\;$10^4$) compared to the spin-relaxation time to pump-down time ratio, the similar pump-down time observed over all detunings (suggesting an inhomogeneous broadened resonance line), and no observed softening of the exponent or \Tone\ saturation~\cite{Fu2006msf} at the measured fields below 3T with longer \Tone (Fig~\ref{fig:Bdep}).
For measuring \Tone\ further from resonance, increasingly longer pump-on times and varying integration window times were used to fully initialize the system.
In control measurements, we find that pump-power, pump-on time (App.~\ref{app:transient}) and integration window (App.~\ref{app:window}) do not significantly affect \Tone\ .

\begin{figure}[!t]
    \centering
     \includegraphics[width=0.976\linewidth]{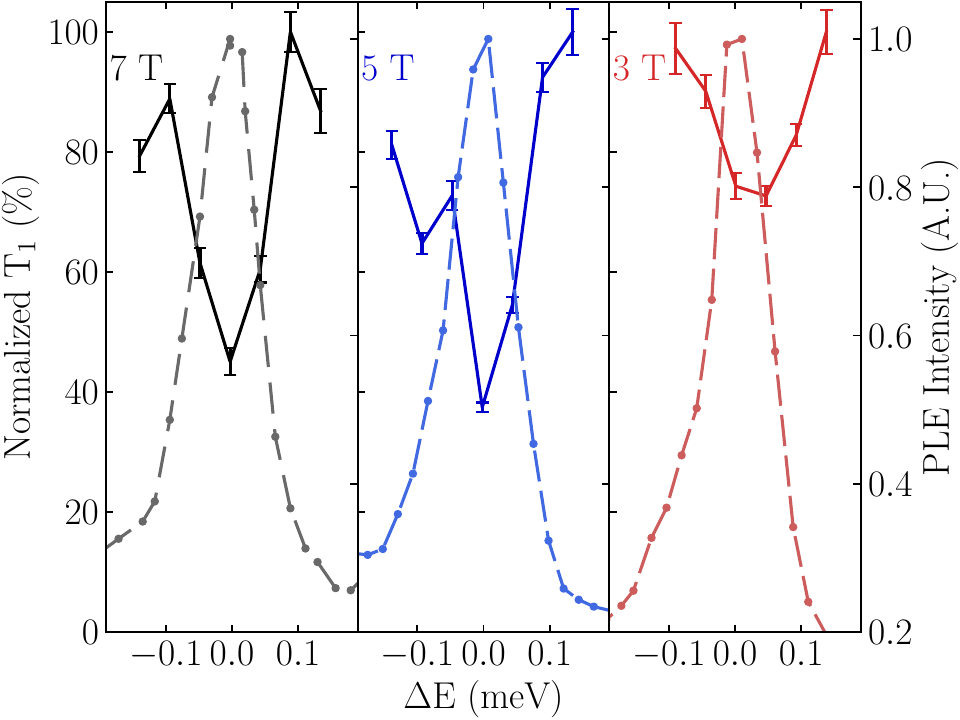}
    \caption{Normalized \Tone\ measurements at 1.8\,K, with varying field, and excitation energy in the Faraday geometry (solid line, left axis) and the corresponding PLE spectra (dashed line, right axis). The measurements were taken on a different spot on the sample than the ones on Fig.~\ref{fig:EnDep}(a). The error bars correspond to one standard deviation of the \Tone\ fitting error. The maximum \Tone\ observed are 1.55\,ms, 6.69\,ms, 53.6\,ms for 7\,T, 5\,T, and 3\,T respectively, and are equivalent to 100\% of the normalized \Tone.}
    \label{fig:BnEnDep}
\end{figure}

\begin{figure}[ht]
    \centering
     \includegraphics[width=0.98\linewidth]{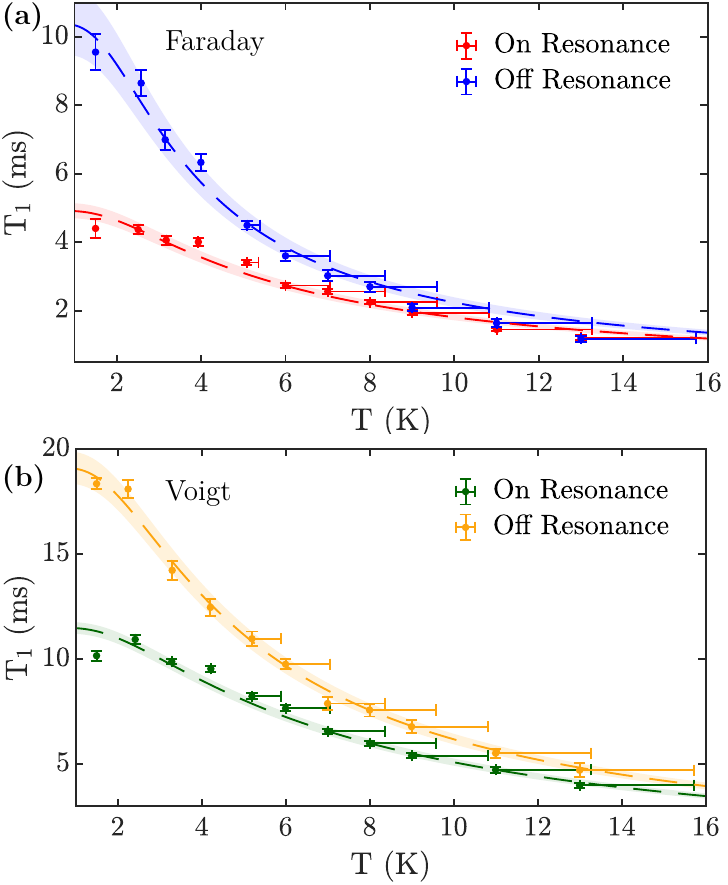}
    \caption{Spin-relaxation time as a function of temperature at $B = 5$\,T. The error bars in \Tone ~correspond to one standard deviation of the \Tone\ fitting error.
    The increasing uncertainty in temperature rises from a systematic underestimation of the temperature due to the distance and lack of thermal contact between the temperature sensor and the sample.
    The dashed lines are least-square fits to the function $(\Gamma_{\downarrow\uparrow} F_{\mathrm{ph}}(T)+\Gamma_0)^{-1}$. 
    The shaded areas around each fit depict the model function with $\Gamma_{\downarrow\uparrow} = \Gamma_{\downarrow\uparrow, \mathrm{fit}} \pm \Gamma_{\downarrow\uparrow, \mathrm{fit, err}}$.
    (a) Faraday,
    $\Gamma_{\downarrow\uparrow} = 0.1647 \pm 0.0091$\,ms$^{-1}$,
    $\Gamma_{0,\mathrm{on}} = 0.0386 \pm 0.0144$\,ms$^{-1}$, $\Gamma_{0,\mathrm{off}} = -0.0685 \pm 0.0108$\,ms$^{-1}$
    and (b) Voigt, 
    $\Gamma_{\downarrow\uparrow} = 0.0512 \pm 0.0021$\,ms$^{-1}$, 
    $\Gamma_{0,\mathrm{on}} = 0.0357 \pm 0.0035$\,ms$^{-1}$, 
    $\Gamma_{0,\mathrm{off}} = 0.0011 \pm 0.0027$\,ms$^{-1}$.
    }
    \label{fig:Tdep}
\end{figure}

We further investigate the size of the energy dependence of \Tone\ as a function of field. Figure~\ref{fig:BnEnDep} depicts the change of the excitation energy dependence with varying magnetic field in the Faraday geometry. 
We observe that the \Tone\ variation does not exceed a factor of 1.25 for low fields (3\,T), but can vary by more than a factor of two at higher fields (5\,T, 7\,T). 
Hence, the choice of excitation energy can impact the magnetic field dependence shown in Fig.~\ref{fig:Bdep}. 
For Faraday geometry, the magnetic field dependence deviates from the theoretically predicted behavior at the higher fields ($B\geq5$\,T) where the energy-dependent deviation is largest. 
However, we note that the softening of the exponent at high fields would be even greater if the magnetic-field dependence had been measured in the off-resonance condition.

The higher spin relaxation at larger \DoX\ intensity (indicating higher donor density) suggests an additional relaxation mechanism based on donor-donor interactions.
The origin of this relaxation mechanism is unknown at this time.
We can rule out exchange and dipolar donor-donor interactions. 
The Bohr radius of the electron donor can be estimated to be $\sim$1.5\,nm~\cite{Meyer:2004tt}.
Donor densities on the order of $10^{16}$\,cm$^{-3}$ yield an average distance between donors of $\sim$30\,nm, meaning that exchange interactions would have little to no effect on the the ensemble longitudinal spin-relaxation process. 
Dipolar interaction on the other hand would yield a flip-flop rate of approximately 10 - 1000\,Hz, comparable to the the experimentally observed relaxation rate. 
However, the hyperfine interaction of the donor with the lattice $^{67}$Zn induces inhomogeneity of tens of MHz in the Zeeman energies~\cite{linpeng2018cps}.
Due to energy conservation, this hyperfine interaction should effectively block dipolar donor-donor flip-flops in the absence of an additional energy-conserving mechanism. 

The temperature dependence of \Tone\ at 5\,T is shown in Fig.~\ref{fig:Tdep}. 
The measurements were taken in the two magnetic field orientations and at two excitation energies which we label ``on-resonance" and ``off-resonance". 
The on-resonant measurements are performed at the wavelength near the minimum \Tone. 
For the off-resonant measurements, we excite 44\,μeV (10.6\,GHz) and 118\,μeV (28.5\,GHz) negatively detuned from the resonance condition for the Faraday and Voigt geometries, respectively. 
At 1.5\,K, \Tone\ values between the two excitation resonance conditions differ by approximately a factor of 2. 
At high temperatures ($T>10$\,K), the on- and off-resonance relaxation times converge. 

We are able to obtain reasonable agreement to a simple relaxation model in which the total spin relaxation is proportional to a sum of a phonon-dependent and a constant term; $(T_1)^{-1} = \Gamma_{\downarrow\uparrow}F_{\mathrm{ph}}(T) + \Gamma_0$, with phonon factor $F_{\mathrm{ph}}(T) = 2 N_{\mathrm{ph}}(T) +1$ (compare with Eq.~\ref{eq:T1}). The fit was performed with a common $\Gamma_{\downarrow\uparrow}$ for both on- and off-resonance datasets, and different $\Gamma_{0,\mathrm{on}}$ and $\Gamma_{0,\mathrm{off}}$.

As shown in Fig.~\ref{fig:Tdep}, this simple temperature dependence model describes both on- and off-resonance datasets. 
In both geometries we find a significantly larger $\Gamma_0$ for on-resonance than off-resonance, consistent with the excitation energy dependence. 
In the Faraday off-resonance case, the fit in fact produces a negative $\Gamma_0$, indicating that the additional relaxation mechanism cannot solely be explained by a simple constant. 
This could indicate a suppression of the phonon-induced spin-relaxation rate when detuned from the ensemble resonance, however the simplicity of the model (which does not include a temperature dependence for $\Gamma_0$ or the effect of temperature on the homogeneous exciton linewidth) prevents a firm conclusion. 
Thus, the origin of the additional relaxation (or stabilization) mechanism is a subject for future study.   

\section{Concluding remarks}
\label{sec:conclusion}
In summary we have demonstrated long longitudinal relaxations times of up to 480 ms for shallow donors in ZnO. 
The measured \Tone\ is approximately three orders of magnitude longer than prior work in other direct bandgap materials (GaAs, CdTe and InP~\cite{ref:linpeng2016lsr}) and stems directly from ZnO's small spin-orbit coupling. Quantitatively we find good agreement of the experimental results with a single-phonon relaxation mechanism. The small longitudinal spin relaxation supports the promise of donors in direct band-gap II-VI semiconductors in which isotope purification is possible to enable long spin coherence times. 

\section{Acknowledgements}
\label{sec:ack}
We thank M. Glazov for fruitful discussions. This material is based upon work supported by the National Science Foundation under Grant No. 1820614. M.V.D. acknowledges the financial support from the Basis Foundation for the Advancement of Theoretical Physics and Mathematics.
\appendix
\section{Satellite band transitions}
\label{app:tes_spectrum}
\begin{figure}[ht]
    \centering
     \includegraphics[width=0.976\linewidth]{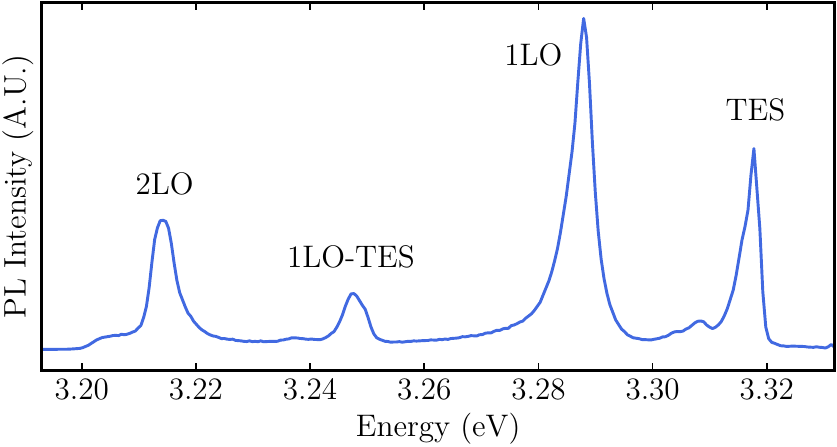}
    \caption{Satellite band transitions in Faraday geometry at 5\,T, and 1.5\,K with resonant excitation of the Ga \DoX ~line.
}
    \label{fig:satelliteband}
\end{figure}

When we collect off-resonance photoluminescence, we collect a wide band of energies, ranging from 3.20\,eV to roughly 3.32\,eV.
Within this broad band, we have identified the lines at 3.318\,eV, 3.288\,eV, 3.247\,eV and 3.214\,eV as the TES, 1LO, 1LO-TES and 2LO transitions for the Ga donor respectively~\cite{wagner2011bez} (Fig.~\ref{fig:satelliteband}).

\section{Low magnetic field optical pumping}
\label{app:low_field_OP}
\begin{figure}[ht]
    \centering
     \includegraphics[width=0.976\linewidth]{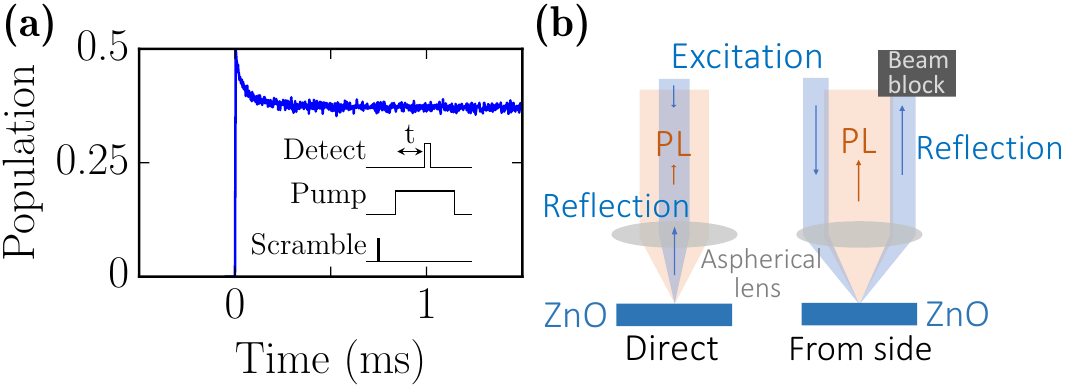}
    \caption{(a) Optical pumping of electrons from the $\ket{\downarrow}$ state to the $\ket{\uparrow}$ state (exciting the $\sigma^+$ transition) in Faraday geometry, 0.3\,T, 1.5\,K . A scrambling pulse is used to initialize both neutral donor electron states to 50\,\%.
    (b) Optical paths of the excitation beam and emitted photoluminescence in the side-excitation scheme.
}
    \label{fig:op0p3T}
\end{figure}

In Sec.~\ref{sec:measurement}, we discussed two ways to verify optical pumping. 
In the Faraday geometry at low field, the optical pumping contrast becomes too low to be detected via the satellite band transitions due to laser background in the corresponding collection energy region. 
Instead, the excitation laser beam was offset on the focusing lens. 
This side-excitation scheme allows for the reflected excitation beam and the emitted photoluminescence to be spatially filtered as depicted in Fig.~\ref{fig:op0p3T}(b) and resonant photoluminescence to be collected.

Figure~\ref{fig:op0p3T}(a) depicts an optical pumping trace at 0.3\,T. 
At such low magnetic fields, \Tone\ is expected to be very long and hence the wait time between pump-on pulses would deem the experiment very slow. 
To speed up the measurement, we utilize a short scrambling pulse at 3.45\,eV to initialize the two \Do\ electron spin states to 50\,\%.

\section{Dependence of \texorpdfstring{T\textsubscript{1}}{T 1} on pump-on time and excitation power}
\label{app:transient}

For measuring \Tone\ as a function of the excitation energy (see Sec.~\ref{sec:exp_energy_temp}), the pump-on time was varied to achieve initialization of the ground-state spins (see Sec.~\ref{sec:measurement}). 
It is possible that the observed excitation energy dependence of \Tone\ (see Sec.~\ref{sec:exp_energy_temp}) might be due to a variation in pump-on time. Additionally, while the nominal excitation power was kept constant during these measurements, 
it is interesting to test whether \Tone\ displays a dependence on excitation power in order to understand limiting factors for \Tone. 
Thus, we performed experiments to gauge the influence of different pump-on times and nominal excitation powers on \Tone.
Performing such experiments similar to the approach presented in Sec.~\ref{sec:measurement} is challenging, for example, because at very high excitation powers it is not possible to detect the start population due to very fast optical pumping.
This problem can be mitigated by conducting the experiment using two lasers (pump-probe experiment).

\begin{figure}[!t]
    \centering
     \includegraphics[width=0.976\linewidth]{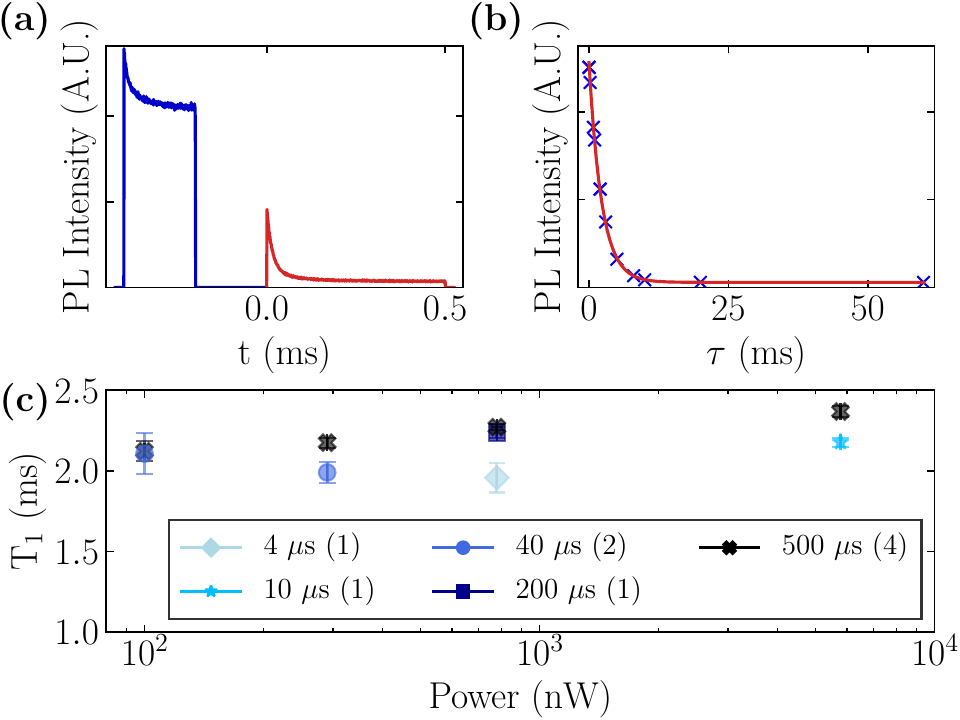}
    \caption{(a) Optical pumping curve using a pump and a probe laser in the Faraday geometry at 5\,T and 1.9\,K. The excitation energy was chosen to be close to the maximum of the ensemble resonance. The inset shows the color-coded spectral position of the pump and probe levels in the energy level diagram. 
    (b) Spin-relaxation trace in the same condition. The red curve is the fit curve from which we extract \Tone. The inset shows the OP and \Tone\ pump-probe measurement schemes. The integration window used is highlighted with gray color both in the inset and in (a).
    (c) \Tone\ pump-probe measurements conducted with various pump excitation powers and pump-on times (with constant probe conditions). The horizontal axis depicts the pump powers, while the shape and shade of each point represent the pump-on time. The number of points of each color is displayed between parentheses in the legend.
    }
    \label{fig:powerDep}
\end{figure}

In Faraday geometry, the pump laser will be set resonant with the $\sigma^+$ transition of a specific sub-ensemble, and for this sub-ensemble, the $\ket{\downarrow}$ population is transferred to the $\ket{\uparrow}$ population.
The specific amount of population that is transferred depends on the pump-on time and/or excitation power of the pump laser.
After a delay time $\tau$, the probe laser, resonant with the $\sigma^-$ transition of the sub-ensemble, probes the remaining $\ket{\uparrow}$ population (see Fig.~\ref{fig:powerDep}(a)). 
Plotting this population as a function of $\tau$ can be used to obtain \Tone. We then measure \Tone\ while tuning the pump parameters, allowing the \Tone\ dependence on pump-on time and excitation power to be measured without varying the probing conditions (see Fig.~\ref{fig:powerDep}(b)). 

As shown in Fig.~\ref{fig:powerDep}(c), \Tone\ does not vary more than a factor of 1.2 in dependence on excitation power and/or pump-on time. 
Thus, it is unlikely that the specific parameters chosen for the \Tone\ measurements have a significant influence on the measured value for \Tone. In Fig.~\ref{fig:powerDep}(c), the data points with a pump-on time of 500\,μs resemble closest the conditions used in the main text for determining \Tone\ because complete spin initialization has been achieved.

\section{Dependence of \texorpdfstring{T\textsubscript{1}}{T 1} on optical pumping integration time}
\label{app:window}

When measuring \Tone\ via optical pumping, we observe different pump-down time for different experimental conditions.
Since the goal is to only collect signal from the beginning of the pump-down trace, which is proportional to the population of the state of interest, we need to integrate the signal in the smallest possible time window. 
However, the smaller the integration window (or gate-on time), the less signal we collect, leading to unsatisfactory statistics.
In order to balance the two effects, we choose different gate-on times for each measurement in the main text. 

\begin{figure}[t]
    \centering
     \includegraphics[width=0.976\linewidth]{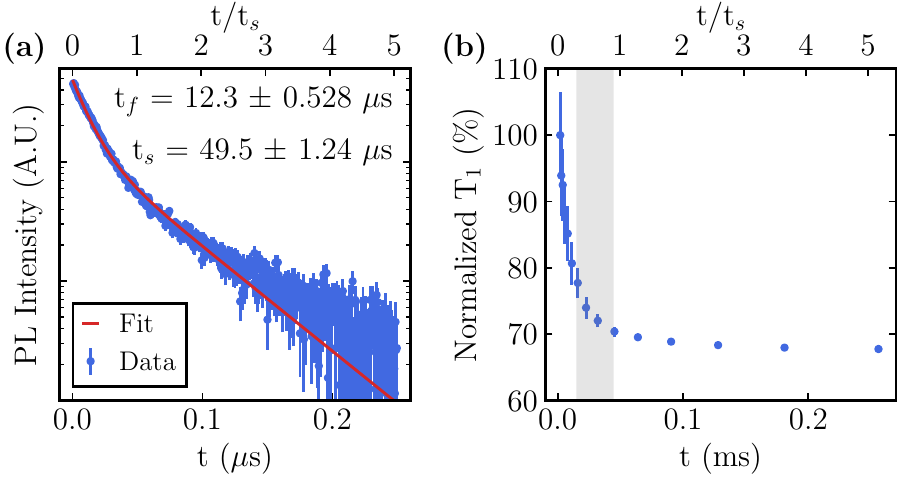}
    \caption{(a) Optical pumping trace using a pump and a probe laser in the Faraday geometry at 5\,T and 1.9\,K, fit with a double exponential model, with fast and slow decays of $t_\mathrm{f}$ and $t_\mathrm{s}$ respectively.
    The y-axis is in logarithmic scale.
    (b) Normalized \Tone\ from a single pump-probe experiment, as a function of the window integration time.
    The gray area depicts the the experimental choices on gate-on time, 0.3$\times t_\mathrm{s}$ to 0.9$\times t_\mathrm{s}$.
}
    \label{fig:windowDep}
\end{figure}

As shown in Fig.~\ref{fig:windowDep}(a), the OP curve can be sufficiently described by a double exponential decay model. The fast decay time of the OP is roughly up to one order of magnitude shorter than the slow decay time. For the excitation energy dependence experiments, the gate-on time remained between 0.3 to 0.9 times the slow component of the OP curves. 
To investigate the effect of choosing such a wide range of gate-on times we utilize a two laser pump-probe experiment (see App.~\ref{app:transient}). Fig.~\ref{fig:windowDep}(b) shows the normalized \Tone\ as a function of gate-on time normalized to the slow decay time. We observe that for normalized gate-on times of 0.3 and 0.9 the variation of \Tone\ is negligible ($\sim$8\,\%). It is interesting to note that for much smaller normalized integration windows, \Tone\ can vary as much as 20\,\%. 
However, we do not probe such short gate-on times due to the low count rate.
Overall, we conclude that the integration window choice does not significantly change the observed \Tone\ for all experimental data.

\section{Faraday geometry g-factor}
\label{app:gfac}

\begin{figure}[t]
    \centering
    \begin{tikzpicture}
        \draw (0, 0) node[inner sep=0] {\includegraphics[width=\linewidth]{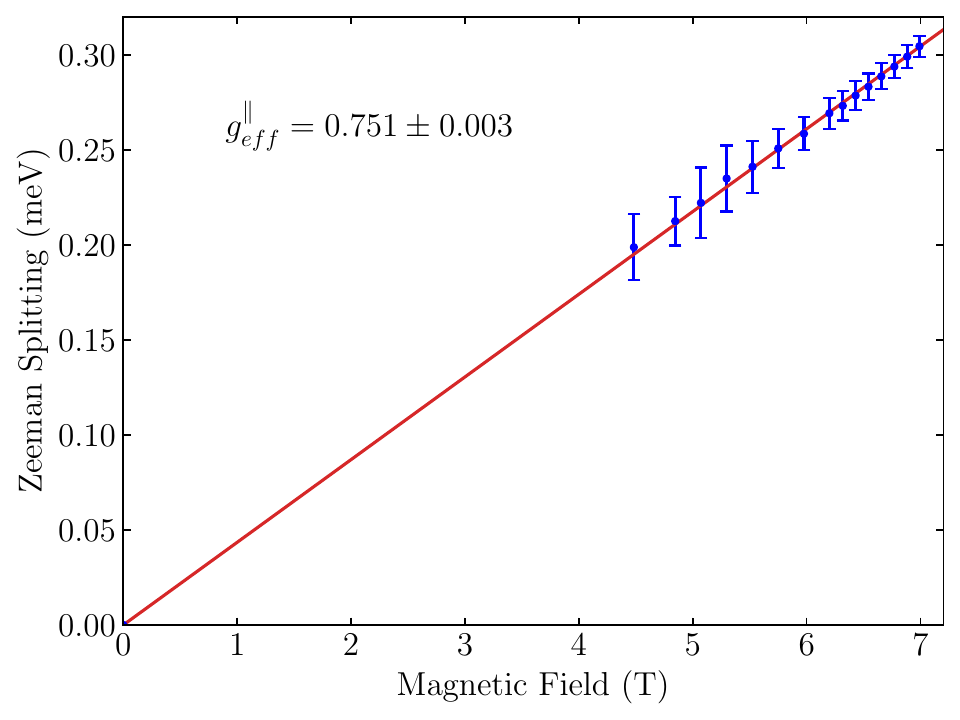}};
    \end{tikzpicture}
    \caption{Zeeman splitting of the Ga donor lines as a function of magnetic field at 5.2\,K in the Faraday geometry. Each point is obtained via Voigt profile fits on PL spectra under 3.45\,eV excitation at different fields. The error bars depict the standard deviation error of the Voigt profile fits.
}
    \label{fig:gfact}
\end{figure}

The g-factors for the Ga donors in this sample in the Voigt geometry have been estimated by linear fits of the
electron and hole Zeeman splitting at different fields.
In prior work, we found that  $g^{\perp}_e$ = 1.97 ± 0.01 and $g^{\perp}_h$ = 0.34 ± 0.02~\cite{linpeng2018cps}. 
To determine the hole g-factor in the Faraday geometry, we fit the transition splitting of the $\sigma+$ and $\sigma-$ transitions (Fig.~\ref{fig:gfact}). 
The resulting effective g-factor is the difference between the electron and the hole g-factors. If we assume $g^{\parallel}_e$ = $g^{\perp}_e$ = 1.97, we estimate $g^{\parallel}_h$ = -1.22 ± 0.01, which is in agreement with the literature~\cite{Wagner2009gvb}.

\normalem
\bibliography{references.bib}
    
\end{document}